\documentstyle[12pt,aasms4]{article}

\begin{document}

\noindent
{\it Conference Highlights}

\begin{center}

\title{\large \bf The Central Kiloparsec of Starbursts and AGN: \\ The
La Palma Connection\footnote{Conference was held in La Palma, Spain in
May 2001. Proceedings have been edited by J.H.\ Knapen, J.E.\ Beckman, I.\
Shlosman, \& T.J.\ Mahoney and published in the {\it ASP Conference
Series.}, Vol. 249 (2001)}}

\end{center}

\medskip

From May 7 to 11, 2001, the Hotel Hacienda San Jorge in the small resort
town of Los Cancajos on La Palma was filled with astronomers studying
the central kiloparsec regions of starbursts and AGN. La Palma is one of
the smallest of the Spanish Canary Islands and is well known for its
world-class astronomical observatory at the Roque de los Muchachos. The
main telescopes operated there are the Isaac Newton Group's 4.2m William
Herschel Telescope and the Italian 3.5m National Telescope, while Spain
is presently constructing its 10.4m telescope, Grantecan. In spite of a
strong astronomical presence of at least two decades, this was the first
scientific astronomy conference organized completely on La Palma. The
fact that many high-resolution studies of galaxies have been made using
data obtained on La Palma makes the island a fitting ``connection'' to
the field of study. The meeting was attended by 120 astronomers from 15
countries, while a number of people had to be disappointed because the
capacity of the hotel and conference room would not allow more
attendees. The program was full and included nine oral and two poster
sessions, long lunch breaks, which were often used for further
discussions or collaborative work, and a number of social events. Most
of those attending participated in a guided tour of the telescopes at
the Observatory on the Wednesday afternoon.

Recent advances in high-resolution observations, theory and modeling
have focused our attention on the central kiloparsec regions of nearby
disk galaxies, which often show profound starburst and/or nonstellar
(AGN) activity, accompanied by intricate gas and dust morphologies and
kinematics. The origin and evolution of the phenomena occurring in these
central regions, their possible causal interrelationships, links to the
host galaxies, and the role the central regions play in galaxy
evolution, were the main topics under consideration.

The meeting brought together researchers in a variety of specialties,
who could concentrate on a relatively restricted part of the complete
field. This focused the discussions and led to a very productive
exchange of ideas. One of the topics covered in detail was the
connection between an AGN and a starburst, which are present
simultaneously in a sufficient fraction of cases to have stimulated the
hypothesis that if they are not the same phenomenon, at least there
might be a causal relation between them.  There is now convincing
evidence from observations across the spectrum for this, and the meeting
brought to bear many new examples, including beautiful new results
obtained with orbiting X-ray observatories. It is now accepted that AGN
and starburst activity often occur together, and this strengthens the
view that the two phenomena are triggered and fueled by common
mechanisms. No direct causal relationships or clear evolutionary
scenarios between AGN and starbursts have yet emerged, but we are
progressing quickly and can expect to find them if they are in fact
present.

Many detailed studies of the physics of star formation and non-stellar
activity were presented during the meeting, both theoretical and
observational, the latter using tracers at wavelengths ranging from the
X-ray to radio domains. These included investigations of the mechanisms
which form jets in disks, of the torus and narrow- and broad-line
regions in AGN, of the role of magnetic fields in the central
kiloparsec, and of the mechanisms leading to enhanced star formation at
specific locations and times. Important new results, both from
observations and from population synthesis modeling, were presented on
the properties and especially the ages of the stellar clusters found in
starburst regions, either central or circumnuclear.

The relations between an AGN and the properties of its host galaxy were
explored at length, albeit without reaching clear conclusions.  There is
now further evidence for a significant though weak trend for Seyfert
galaxies to be more often barred than non-Seyfert galaxies.  Many
beautiful and detailed numerical and observational studies of the
morphology and kinematics of nuclear bars, rings and spiral arms were
presented, but, perhaps disappointingly, no direct links to AGN fueling
were discovered.  It is possible, and theoretically quite plausible,
that any relation between circumnuclear and host galaxy structure and
the presence or properties of an AGN may be weakened by the tendency of
a central, or circumnuclear, mass concentration to dissipate the bar
which originally induced it.

Following notable recent observational advances, the relationship
between supermassive black holes, which are assumed to drive the AGN,
and their surrounding spheroidal component structures (the whole galaxy
for an elliptical, and the bulge for a spiral) received considerable
attention.  A related topic, the relation between local gas density and
star formation rate, was discussed starting from conclusions based on
its conventional context: galactic disks, and going on to discuss the
implications for circumnuclear zones. A theoretical and observational
attack on this problem would in all probability go a long way to
explaining why there is AGN activity around some supermassive black
holes, but less or none around others.

In general, it was heartening to see real progress being made across the
field thanks notably to a continuing stream of {\it Hubble Space
Telescope} observations, to X-ray results obtained with new orbiting
observatories, but also to imaging and spectroscopy from ground-based
optical/near-infrared telescopes, such as those on La Palma. In
addition, numerical simulations of gas dynamics are progressing quickly,
reaching higher dynamical and spatial resolution and deepening our
insight into a range of physical processes.

Any conference proceedings necessarily provides a snapshot of the state
of a given field, which makes it appear perhaps more static than it
really is. In this field, we do foresee a continuing, even increasing,
high rate of research leading to important breakthroughs both
observationally, due to new instrumental opportunities allowing
observation at higher angular resolution, sensitivity, area coverage,
and wavelength coverage, and theoretically, due to continuing major
refinements in modeling. We trust that the proceedings of the present
conference will, then, offer not only a fixed photograph of the present,
but a useful, even prophetic guide to the future of this exciting and
changing subject.

Financial support for the conference was given by the Excelent\'\i simo
Cabildo Insular de La Palma (Island Government), its Patronato de
Turismo (Tourist Board), and the Isaac Newton Group of Telescopes, for
which we are very grateful.

\medskip

\noindent
{\it Johan H.\ Knapen} 

\noindent
{\it Isaac Newton Group of Telescopes, Spain, and University of
Hertfordshire, U.K.}

\noindent
{\it John E.\ Beckman}

\noindent
{\it Instituto de Astrof\'\i sica de Canarias, Spain}

\noindent
{\it Isaac Shlosman}

\noindent
{\it Joint Institute for Laboratory Astrophysics, and University of
Kentucky}


\noindent
{\it Terry J.\ Mahoney}

\noindent
{\it Instituto de Astrof\'\i sica de Canarias, Spain}

\end{document}